\documentclass[preprint,showpacs,superscriptaddress,nofootinbib]{revtex4}
  \usepackage{graphicx,subfigure}
   \usepackage{color}
 \usepackage{ulem}
  \begin{document}
  \title{A Combined Fit on the Annihilation Corrections in $B_{u,d,s}$
         $\to$ $PP$ Decays Within QCDF}
  \author{Qin Chang}
  \affiliation{Institute of Particle and Nuclear Physics,
              Henan Normal University, Xinxiang 453007, China}
  \affiliation{State Key Laboratory of Theoretical Physics, Institute of Theoretical Physics,
        Chinese Academy of Sciences, China}
  \author{Junfeng Sun}
  \affiliation{Institute of Particle and Nuclear Physics,
              Henan Normal University, Xinxiang 453007, China}
  \author{Yueling Yang}
  \affiliation{Institute of Particle and Nuclear Physics,
              Henan Normal University, Xinxiang 453007, China}
  \author{Xiaonan Li}
  \affiliation{Institute of Particle and Nuclear Physics,
              Henan Normal University, Xinxiang 453007, China}
  \begin{abstract}
  Motivated by the possible large annihilation contributions
  implied by recent CDF and LHCb measurements on nonleptonic
  annihilation $B$-meson decays, and the refined experimental
  measurements on hadronic $B$-meson decays, we study the
  strength of annihilation contributions within QCD
  factorization (QCDF) in this paper.
  With the available measurements of two-body $B_{u,d,s}$ ${\to}$
  ${\pi}{\pi}$, ${\pi}K$, $KK$ decays, a comprehensive
  fit on the phenomenological parameters $X_{A}^{i,f}$
  (or ${\rho}_A^{i,f}$ and ${\phi}_A^{i,f}$)
  which are used to parameterize the endpoint singularity
  in annihilation amplitudes is performed with the statistical
  $\chi^2$ approach. It is found that
  (1) flavor symmetry breaking effects are
  hardly to be distinguished between $X_{A,s}^i$ and
  $X_{A,d}^i$ due to the large experimental errors
  and theoretical uncertainties, where $X_{A,s}^i$ and
  $X_{A,d}^i$ are related to the nonfactorization
  annihilation contributions in $B_{s}$ and $B_{u,d}$
  decay, respectively. So $X_{A,s}^i$ ${\simeq}$ $X_{A,d}^i$
  is a good approximation by now.
  (2) In principle, parameter $X_{A}^f$ which is related to the
  factorization annihilation contributions and
  independent of the initial state can be regarded
  as the same variable for $B_{u,d,s}$ decays.
  (3) Numerically, two solutions are found, one is
  $(\rho_A^i, \phi_A^i[^{\circ}])$ $=$ $(2.98^{+1.12}_{-0.86},-105^{+34}_{-24})$
  and
  $(\rho_A^f, \phi_A^f[^{\circ}])$ $=$ $(1.18^{+0.20}_{-0.23},-40^{+11}_{-8})$,
  the other is
  $(\rho_A^i, \phi_A^i[^{\circ}])$ $=$ $(2.97^{+1.19}_{-0.90},-105^{+32}_{-24})$
  and
  $(\rho_A^f, \phi_A^f[^{\circ}])$ $=$ $(2.80^{+0.25}_{-0.21},165^{+4}_{-3})$.
  Obviously, nonfactorization annihilation parameter $X_{A}^{i}$ is generally
  unequal to factorization annihilation parameter $X_{A}^{f}$,
  which differ from the traditional treatment.
  With the fitted parameters, all results for observables of $B_{u,d,s}$ $\to$
  ${\pi}{\pi}$, ${\pi}K$, $KK$ decays are in good agreement with
  experimental data.
  \end{abstract}
  \maketitle
  With the running of the Large Hadron Collider (LHC),
  many intriguing $B$-meson decays are well measured and some interesting
  phenomena are found by LHCb collaboration in the past years.
  For example, measurements of branching fractions for the pure annihilation 
  $B_{d}$ ${\to}$ $K^{+}K^{-}$ and  $B_{s}$ $\to$ ${\pi}^{+}{\pi}^{-}$ decays
  \cite{LHCbanni}. 
  Their averaged results given by
  Heavy Flavor Averaging Group (HFAG) are \cite{HFAG}
  \begin{equation}
 {\cal B}(B_{d}{\to}K^{+}K^{-})
  =(0.12{\pm}0.05){\times}10^{-6}
  \label{HFAGKK},
  \end{equation}
  \begin{equation}
 {\cal B}(B_{s}{\to} {\pi}^{+}{\pi}^-)
  = (0.73 \pm 0.14) \times 10^{-6}
  \label{HFAGpipi},
  \end{equation}
  which attract much attention recently \cite{xiao1,zhu1,zhu2,chang1}. 

  Theoretically, the branching ratios of pure annihilation nonleptonic $B$
  meson decays are formally $\Lambda_{\rm QCD}/m_b$ power suppressed and
  expected at $10^{-7}$ level, which roughly agrees with the measurements.
  In the framework of QCD factorization (QCDF) \cite{Beneke1},
  the annihilation amplitudes, together with the chirally enhanced
  power corrections and possible large strong phase involved in them,
  play an important role in evaluating the observables of $B$ meson decays.
  However, due to the endpoint singularities, the amplitudes of annihilation
  topologies are hardly to be exactly calculated.
  To estimate the endpoint contributions, phenomenological parameter
  $X_A$ is introduced \cite{Beneke2} as
  \begin{equation}
  \int_0^1 \frac{\!dx}{x}\, \to X_A =
  (1+ \rho_A e^{i \phi_A}) \ln \frac{m_B}{\Lambda_h}
  \label{XA},
  \end{equation}
  where $\Lambda_h$ $=$ $0.5{\rm GeV}$.
  The QCDF approach itself cannot give some information/or constraint on
  parameters $\rho_A$ and $\phi_A$.
  To simplify the calculation, one usually takes the same parameters
  $\rho_A$ and $\phi_A$ for factorizable and nonfactorizable
  annihilation topologies.
  And as a conservative choice, the values of $\rho_A$ $\sim$ $1$ and
  $\phi_A$ $\sim$ $-55^{\circ}$ (named scenario S4) \cite{Beneke2,Cheng1,Cheng2}
  are usually adopted in previous studies on $B_{u,d,s}$ $\to$ $PP$
  decays, which leads to the prediction\footnotemark[1]
  \footnotetext[1]{The second uncertainty comes
  from parameters ${\rho}_{A,H}$ and ${\phi}_{A,H}$.}
  ${\cal B}(B_{d}{\to}K^{+}K^{-})$ $=$
  $(0.10^{+0.03+0.03}_{-0.02-0.03}){\times}10^{-6}$ \cite{Cheng1}
  and ${\cal B}(B_{s}{\to}{\pi}^{+}{\pi}^{-})$ $=$
  $(0.26^{+0.00+0.10}_{-0.00-0.09}) \times 10^{-6}$ \cite{Cheng2}.  
  Clearly, the QCDF's prediction on ${\cal B}(B_{d}{\to}K^{+}K^{-})$
  agrees well with the current measurements considering the experimental
  and theoretical errors, while the QCDF's prediction on 
  ${\cal B}(B_{s}{\to}{\pi}^{+}{\pi}^{-})$ is much smaller than
  the experimental data Eq.(\ref{HFAGpipi}) by about $3{\sigma}$,
  which implies unexpectedly possible large annihilation corrections
  and possible large flavor symmetry breaking effects between
  the annihilation amplitudes of $B_{u,d}$ and $B_{s}$ decays \cite{zhu2,zhu1}.
  Motivated by such mismatch, some works have been done for
  possible solutions and implications.

  Within the QCDF framework, using the asymptotic light-cone
  distribution amplitudes, the building blocks of annihilation
  amplitudes are simplified as \cite{Beneke1,Beneke2}
   \begin{eqnarray}
   A_1^i & \simeq &
   A_2^i   \simeq 2\, \pi \, \alpha_s \Big[
    9\,\Big( X_A^i - 4 + \frac{\pi^2}{3} \Big)
    + r_\chi^{M_1} r_\chi^{M_2} (X_A^{i})^2 \Big]
  \label{a-i-12}, \\
   A_3^i & \simeq & 6\, \pi \, \alpha_s \,
   (r_\chi^{M_1} - r_\chi^{M_2})
   \Big[ (X_A^{i})^2 - 2 X_A^i + \frac{\pi^2}{3} \Big]
  \label{a-i-3}, \\
   A_3^f & \simeq & 6\, \pi \, \alpha_s \,
   (r_\chi^{M_1} + r_\chi^{M_2})
   \Big[ 2\, (X_A^{f})^2 - X_A^f \Big]
   \label{a-f-3},
   \end{eqnarray}
  where the superscripts ``$i$'' and ``$f$'' refer to gluon
  emission from the initial- and final- states, respectively;
  the subscripts ``1'', ``2'' and ``3'' correspond to three
  possible Dirac structures, with ``1'' for $(V-A) \otimes (V-A)$,
  ``2'' for $(V-A) \otimes (V+A)$, and ``3'' for $(S-P) \otimes (S+P)$,
  respectively; $A_{3}^{i}$ is negligible for light final pseudoscalars
  due to $r_{\chi}^{M_{1}}$ ${\simeq}$ $r_{\chi}^{M_{2}}$.
  The explicit expressions of effective annihilation coefficients
  could be found in Ref. \cite{Beneke1,Beneke2}.
  
 For the annihilation parameters $X_{A}$ in Eqs. (\ref{a-i-12}-\ref{a-f-3}), 
 although there are no imperative and a priori reasons for it to be the same in the 
 building blocks $A_{k}^{i,f}$ ($k$ $=$ $1$, $2$, $3$), the simplification
  $X_{A}^{i}$ $=$ $X_{A}^{f}$ $=$ $X_{A}$ is commonly used in many previous
  works of nonleptonic $B$ decays \cite{Beneke1,Beneke2,Cheng2,du1,Cheng1},
  independent of mesons involved and topologies.
  However, the carefully renewed study in Refs. \cite{zhu1,chang1} shows
  that it is hardly to accommodate all available observables of
  charmless $B$ ${\to}$ $PP$ decays simultaneously with the
  universal ${\rho}_{A}$ and ${\phi}_{A}$.
  Recently, a refreshing suggestion was proposed in Refs. \cite{zhu2,zhu1}
  to cope with the parameters $X_{A}$.
  The main points of ``new treatment" could be briefly summarized as follow:
  \begin{itemize}
  \item[(i)] 
  As the superscripts of ${A}^{i,f}_{k}$ correspond to different
  topologies, parameters of $X_{A}^{i}$ and $X_{A}^{f}$ should
  be treated individually.
  \item[(ii)] 
  For the factorizable annihilation topologies,
  the information of initial state has been included in the decay
  constant of $B$ meson and taken outside from the building
  blocks of $A_{k}^{f}$. Only the wave functions of final states
  are involved in the convolution integral of subamplitudes.
  Additionally, the same asymptotic light cone distribution amplitude
  is commonly applied to the final pseudoscalar and vector mesons.
  So, the parameter $X^{f}_A$ should be universal for factorizable
  annihilation amplitudes of both $B_s$ and $B_{u,d}$ nonleptonic decays.
  \item[(iii)]  
  For the nonfactorizable annihilation topologies,
  the initial $B$ meson entangles with the final states via
  gluon exchange. The wave functions of all participating hadrons,
  including the initial $B$ meson, are involved in the convolution
  integral of subamplitudes.
  Hence, the parameter $X_{A}^{i}$ might be different from
  the parameter $X^{f}_A$ generally.
  Moreover, due to the mass relationship $m_{u}$ ${\simeq}$
  $m_{d}$ ${\neq}$ $m_{s}$ resulting in the $SU(3)$ flavor
  symmetry breaking, it is usually assumed that the momentum
  fraction of the valence $s$ quark in $B_{s}$ meson should be
  larger than that of the spectator $u$, $d$ quark in $B_{u,d}$
  meson.
  The flavor symmetry breaking effects might be embodied in
  parameter $X_{A}^{i}$, i.e. two parameters, $X_{A,d}^{i}$
  and $X_{A,s}^{i}$, should be introduced for nonfactorizable
  annihilation topologies of $B_{u,d}$ and $B_{s}$ meson decay,
  respectively, while the isospin symmetry holds approximately.
  Generally, it is not required that $X_{A,d}^{i}$ must be
  equal or unequal to $X_{A,s}^{i}$, i.e., $X_{A,d}^{i}$ and
  $X_{A,s}^{i}$ are independent variables.
  \end{itemize}
  
  With this assumption, authors of Refs.\cite{zhu2,zhu1} reanalyzed 
  $B_{u,d,s}$ ${\to}$ ${\pi}{\pi}$, ${\pi}K$, $KK$ decays without
  considering theoretical uncertainties and found
  that the experimental data on ${\cal B}(B_{s}{\to}{\pi}^{+}{\pi}^{-})$
  in Eq.(\ref{HFAGpipi}) could be explained with large ${\rho}_{A,s}^{i}$
  ${\sim}$ $3$.
  Compared with ${\rho}_{A}$ ${\sim}$ $1$ in \cite{Beneke2,Cheng1} 
  for ${\cal B}(B_{d}{\to}K^{+}K^{-})$, it seems to imply unexpectedly
  large flavor symmetry breaking effects, then the predictive power of
  QCD will be rather limited. 
  Thanks to the large experimental errors,  ${\cal B}(B_{d}{\to}K^{+}K^{-})$
  can be fitted within a large 
  range of $({\rho}_{A,d}^{i},{\phi}_{A,d}^{i})$ including ${\rho}_{A,d}^{i}$ 
  ${\sim}$ $3$ \cite{zhu1,chang2}. Therefore, flavor symmetry might be 
  restored as both aforementioned decays could be accommodated by a 
  common set of $({\rho}_{A}^{i},{\phi}_{A}^{i})$.  
  It is interesting and essential to systematically evaluate
  the exact strength of annihilation contribution and further
  test the aforementioned points, especially the flavor asymmetry effects.
  
  As it is well known, additional phenomenological parameters $X_{H}$
  (or $\rho_{H}$ and $\phi_{H}$), like to Eq.(\ref{XA}), were introduced
  to regulate the endpoint singularity in the hard spectator scattering (HSS)
  corrections involving the twist-3 light cone distribution amplitudes of
  light final states \cite{Beneke1,Beneke2,Cheng2,du1,Cheng1}. 
  The phenomenological importance of HSS corrections to the 
  color-suppressed tree contributions which are enhanced by the
  large Wilson coefficient $C_{1}$ has already been recognized by
  Refs.\cite{Cheng1,pipipuz,chang2}
  in explicating the current experimental measurements on 
  ${\Delta}A_{CP}$ $=$ $A_{CP}(B^{+}{\to}K^{+}{\pi}^{0})$ $-$ 
  $A_{CP}(B^{0}{\to}K^{+}{\pi}^{-})$ and $R_{00}^{{\pi}{\pi}}$ $=$ 
  $2{\cal B}(B^{0}{\to}{\pi}^{0}{\pi}^{0})/{\cal B}(B^{0}{\to}{\pi}^{+}{\pi}^{-})$.
  Because the $B$ wave functions are also involved
  in the HSS convolution integral, the flavor symmetry breaking 
  effects might be also embodied in parameter $X_{H}$.

  Following the ansatz in Ref. \cite{zhu2,zhu1},
  we preform a global fit on the annihilation parameters
  combining available experimental data on $B_{u,d,s}$
  ${\to}$ ${\pi}{\pi}$, ${\pi}K$, $KK$ decays with 
  a statistical $\chi^2$ analysis.  
  Based on our previous analysis \cite{chang2}, the approximation, 
  ($\rho_{H,d}$, $\phi_{H,d}$) $=$ ($\rho_{A,d}^{i}$, $\phi_{A,d}^{i}$),
  is acceptable by current measurements on $B_{u,d}$ decays 
  (see scenario III in Ref. \cite{chang2} for details),
  which lessens effectively the unknown variables.
  Hence, the approximation $X_{H}$ $=$ $X_{A}^{i}$ is assumed for 
  $B_{u,d,s}$ decays in the following analysis.
  The detailed explanation on the fitting approach could be
  found in the Appendix C of Ref. \cite{chang2}.
  The values of input parameters used in our evaluations
  are summarized in Table \ref{inputs}. 

 \begin{table}[ht]
 \caption{The values of input parameters:
 CKM matrix elements, pole and running quark masses,
 decay constants, form factors and Gegenbauer moments.}
 \label{inputs}
 \begin{ruledtabular}
 \begin{tabular}{l}
 $\bar{\rho}$ $=$ $0.1489^{+0.0158}_{-0.0084}$, \quad
 $\bar{\eta}$ $=$ $0.342^{+0.013}_{-0.011}$, \quad
 $A$ $=$ $0.813^{+0.015}_{-0.027}$, \quad
 $\lambda$ $=$ $0.22551^{+0.00068}_{-0.00035}$ \cite{CKMfitter}
 \\ \hline
 $m_c$ $=$ $1.67 \pm 0.07$ GeV, \quad
 $m_b$ $=$ $4.78 \pm 0.06$ GeV, \quad
 $m_t$ $=$ $173.21 \pm 0.87$ GeV, \\
 $\frac{\bar{m}_s(\mu)}{\bar{m}_{u,d}(\mu)}$ $=$ $27.5 \pm 1.0$, \quad
 $\bar{m}_{s}(2\,{\rm GeV})$ $=$ $95 \pm 5$ MeV, \quad
 $\bar{m}_{b}(\bar{m}_{b})$ $=$ $4.18 \pm 0.03$ GeV \cite{PDG14}
 \\ \hline
 $f_{B_{d}}$ $=$ $(190.6 \pm 4.7)$ MeV, \quad
 $f_{B_{s}}$ $=$ $(227.6 \pm 5.0)$ MeV, \cite{DecayCon} \\
 $f_{\pi}$ $=$ $(130.41 \pm 0.20)$ MeV, \quad
 $f_{K}$ $=$ $(156.2 \pm 0.7)$  MeV \cite{PDG14}
 \\ \hline
 $F^{B \to \pi}_{0}(0)$ $=$ $0.258 \pm 0.031$, \quad
 $F^{B \to K}_{0}(0)$ $=$ $0.331 \pm 0.041$, \quad
 $F^{B_s \to K}_{0}(0)$ $=$ $0.23 \pm 0.06$, \cite{BallZwicky}
 \\
 $a_{1}^{\pi}$ $=$ $0$, \quad
 $a_2^{\pi}(2\,{\rm GeV})$ $=$ $0.17$, \quad
 $a_{1}^{K}(2\,{\rm GeV})$ $=$ $0.05$, \quad
 $a_{2}^{K}(2\,{\rm GeV})$ $=$ $0.17$ \cite{BallG}
 \end{tabular}
 \end{ruledtabular}
 \end{table}
 \begin{figure}[ht]
 \subfigure[]{\includegraphics[width=6cm]{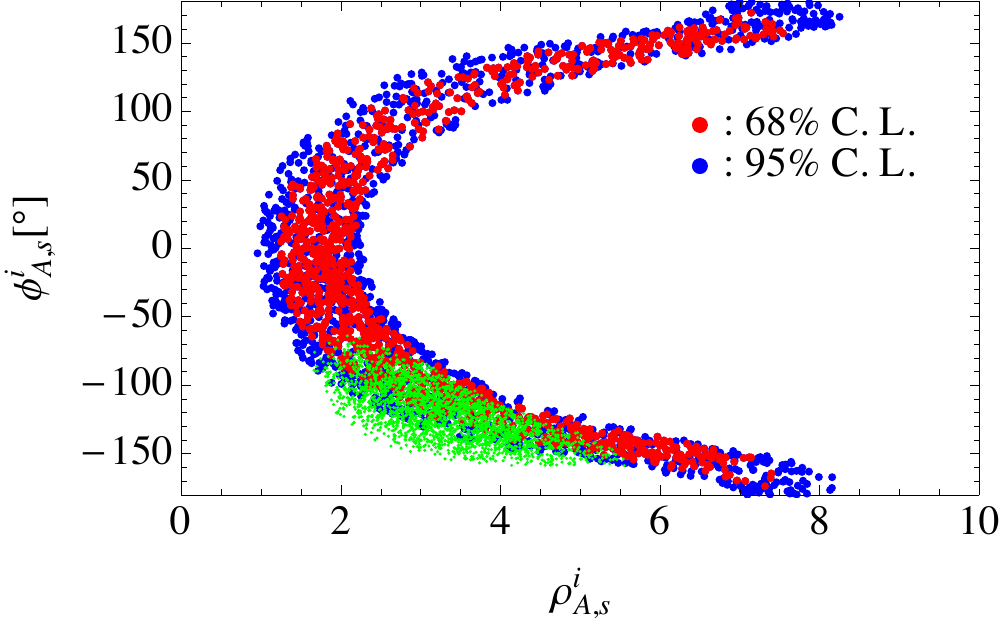}}
 \qquad
 \subfigure[]{\includegraphics[width=6cm]{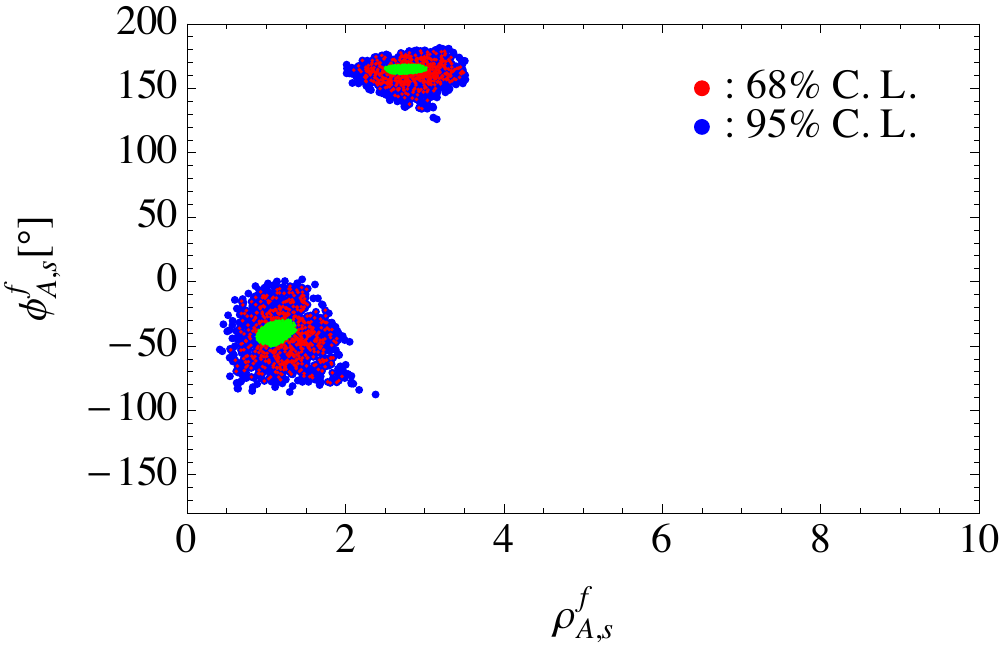}}
 \caption{The allowed regions of annihilation parameters
 at 68\% C.L. and 95\% C.L. for $B_{s}$ decays are
 shown by red and blue points in the planes
 $(\rho_{A,s}^{i,f},\phi_{A,s}^{i,f})$, respectively.
 The green pointed regions are the fitted results of
 $(\rho_{A,d}^{i,f},\phi_{A,d}^{i,f})$ at 68\% C.L.
 for $B_{u,d}$ decays. See text for detail explanation.}
 \label{Bsfit}
 \end{figure}

 Firstly, to clarify the flavor symmetry breaking effects on
 parameters $X_{A}^{i,f}$, we perform a fit on the annihilation
 parameters $(\rho_A^{i,f},\phi_A^{i,f})$ for $B_{u,d}$
 and $B_s$ decays, respectively. 
 For parameters of $(\rho_{A,s}^{i,f},\phi_{A,s}^{i,f})$,
 the constraints come from observables of the $\bar{B}_s$ $\to$
 $\pi^- K^+$, $\pi^+ \pi^-$, $K^+ K^-$ decays.
 The fitted results are shown in Fig.\ref{Bsfit}.
 For parameters of $(\rho_{A,d}^{i,f},\phi_{A,d}^{i,f})$,
 they have been fitted with the constraints from $B_{u,d}$
 $\to$ $\pi K$, $\pi \pi$, $KK$ decays, especially, focusing
 on the so-called ``$\pi K$'' and ``$\pi \pi$'' puzzles (see
 Ref. \cite{chang2} for the details).
 Their allowed regions (green points) at 68\% C.L. are also
 shown in Fig.\ref{Bsfit} for a comparison with
 $(\rho_{A,s}^{i,f},\phi_{A,s}^{i,f})$.
 
 From Fig.\ref{Bsfit}(a), it is seen clearly that 
 (1) the region of $(\rho_{A,s}^{i},\phi_{A,s}^{i})$ cannot
 be seriously constrained by now, because the current measurements
 on $B_s$ $\to$ $\pi^- K^+$, $\pi^+ \pi^-$, $K^+ K^-$ decays
 are not accurate enough and the theoretical uncertainties are
 also still large.  
 Moreover, a relatively large $\rho_{A,s}^{i}$ ${\sim}$ $3$
 with $\phi_{A,s}^{i}$ ${\sim}$ ${\pm}100^{\circ}$ in $B_s$ system 
 suggested by recent studies \cite{zhu2,zhu1,chang1}
 is allowed. 
 (2) The conventional choice of ${\rho}_{A,d}^{i}$ ${\sim}$ $1$ and 
 ${\phi}_{A,d}^{i}$ ${\sim}$ $-55^{\circ}$ \cite{Beneke2,Cheng1,Cheng2}
 is ruled out, because the assumption $X_{H}$ $=$ $X_{A}^{i}$
 is used in our study to enhance the magnitude and the strong phase of the 
 color-suppressed tree amplitude $C$ via spectator interactions and to solve 
 both ``$\pi K$''  and ``$\pi \pi$'' puzzles \cite{chang2}.
 Besides, a relatively large ${\rho}_{A,d}^{i}$ ${\sim}$ $3$ with ${\phi}_{A,d}^{i}$ 
 ${\sim}$ $−100^{\circ}$ is allowed by ${\cal B}(B_{d}{\to}K^{+}K^{-})$ 
 which has large experimental error and theoretical uncertainties until now,
 and is also consistent with Fig.7(a) of Ref.\cite{1409.3252} for $B_{d}$ ${\to}$ 
 $K^{+}K^{-}$ decays using the similar statistical fit approach with 
 parameters $X_{A}^{i}$ $=$ $X_{A}^{f}$.
 (3) The allowed regions of $(\rho_{A,d}^{i},\phi_{A,d}^{i})$ can
 still overlap with the ones of $(\rho_{A,s}^{i},\phi_{A,s}^{i})$
 in part, around ($3,-100^{\circ}$), which implies 
 that the treatment $X_{A,s}^i$ $\neq$ $X_{A,d}^i$ from flavor 
 symmetry breaking effects in Ref. \cite{zhu2,zhu1}
 is not absolutely sure, at least not necessary with current
 experimental and theoretical precision.
 
 From Fig.\ref{Bsfit}(b), it is seen clearly that (1)
 there are two allowed solutions for parameters of both
 $(\rho_{A,d}^{f},\phi_{A,d}^{f})$ and
 $(\rho_{A,s}^{f},\phi_{A,s}^{f})$.
 Besides the commonly used value $\rho_{A}^{f}$ ${\sim}$ $1$
 \cite{Beneke1,Beneke2,Cheng2,du1,Cheng1}, there is another
 best-fit value $\rho_{A}^{f}$ ${\sim}$ $2.5$.
 (2) It is interesting that the allowed regions for
 $(\rho_{A,d}^{f},\phi_{A,d}^{f})$ overlap entirely with
 those for $(\rho_{A,s}^{f},\phi_{A,s}^{f})$, which confirms
 the suggestion \cite{zhu2} that $X_A^f$ (or ${\rho}_{A}^{f}$,
 ${\phi}_{A}^{f}$) is universal for $B_{u,d}$ and $B_s$ system.
 
 Moreover, comparing Fig.\ref{Bsfit} (a) with (b), it is seen that 
 (1) 
 generally, the allowed region of $(\rho_{A}^{f},\phi_{A}^{f})$
 is different from that of $(\rho_{A}^{i},\phi_{A}^{i})$, and 
 $X_{A}^i$ is not always required to be equal to $X_{A}^f$. 
 So, the ``new treatment" on parameters $X_{A}$ according 
 to either factorizable or nonfactorizable annihilation topologies
 may be reasonable and appropriate for $B_{u,d,s}$ decays.
 (2) The flavor symmetry breaking effects on parameters $X_A^{i,f}$
 could be very small even negligible under the existing circumstances 
 with less available experimental constraints from $B_{s}$ decays.
 
 \begin{figure}[ht]
 \subfigure[]{\includegraphics[width=6cm]{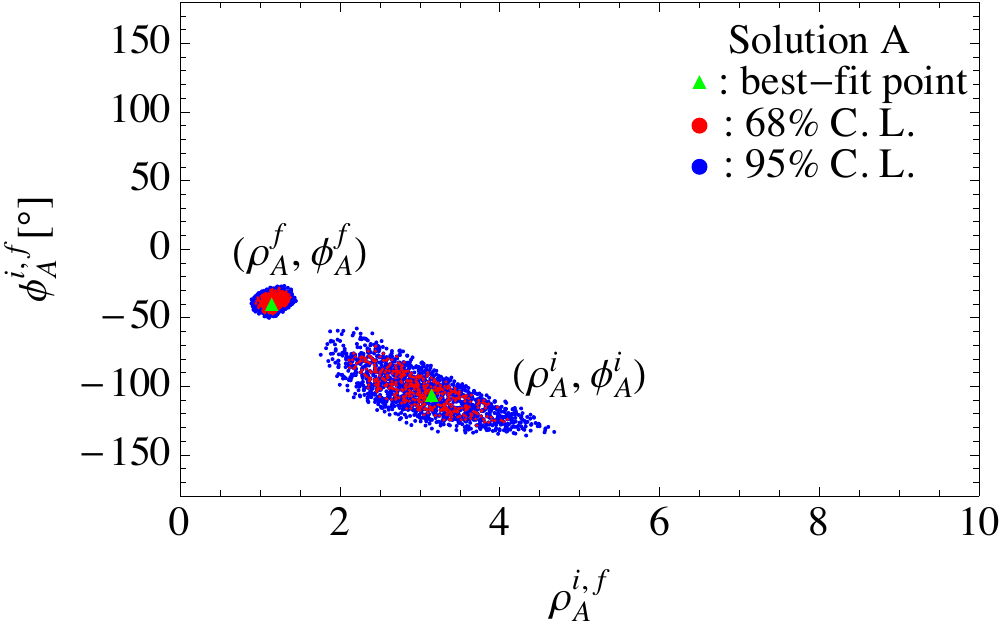}}
 \qquad
 \subfigure[]{\includegraphics[width=6cm]{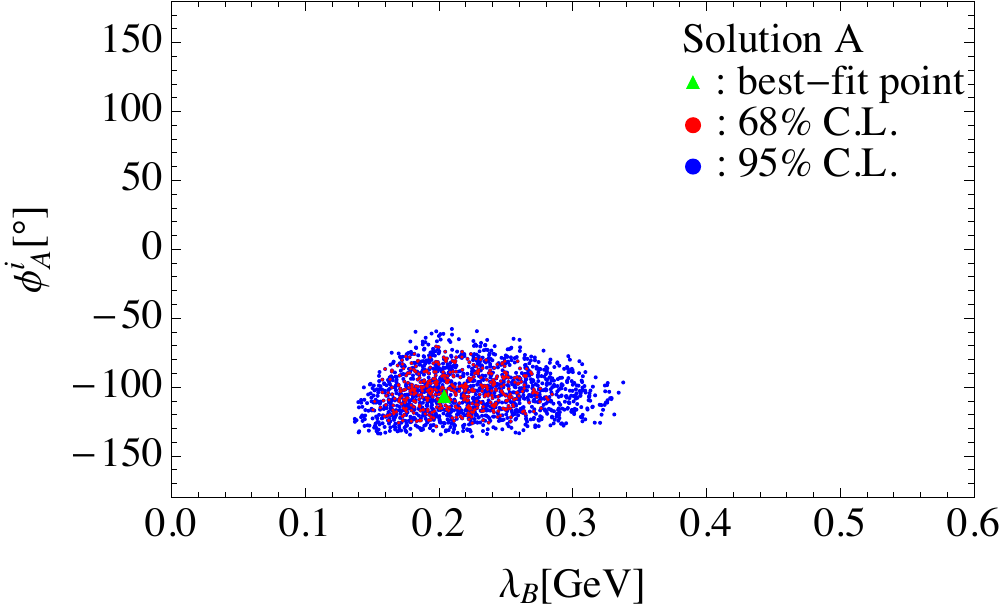}}\\
 \subfigure[]{\includegraphics[width=6cm]{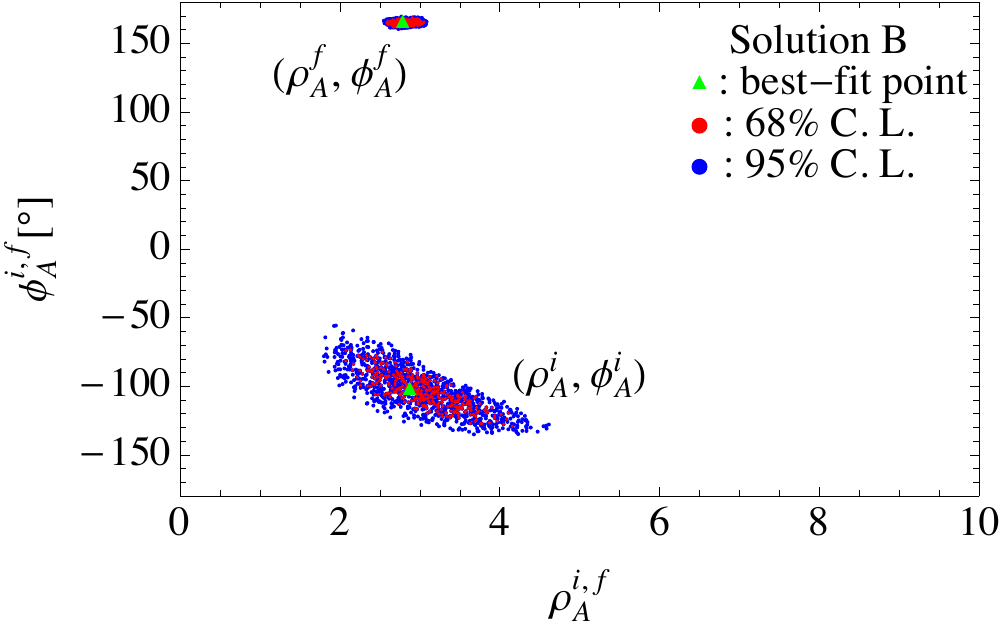}}
 \qquad
 \subfigure[]{\includegraphics[width=6cm]{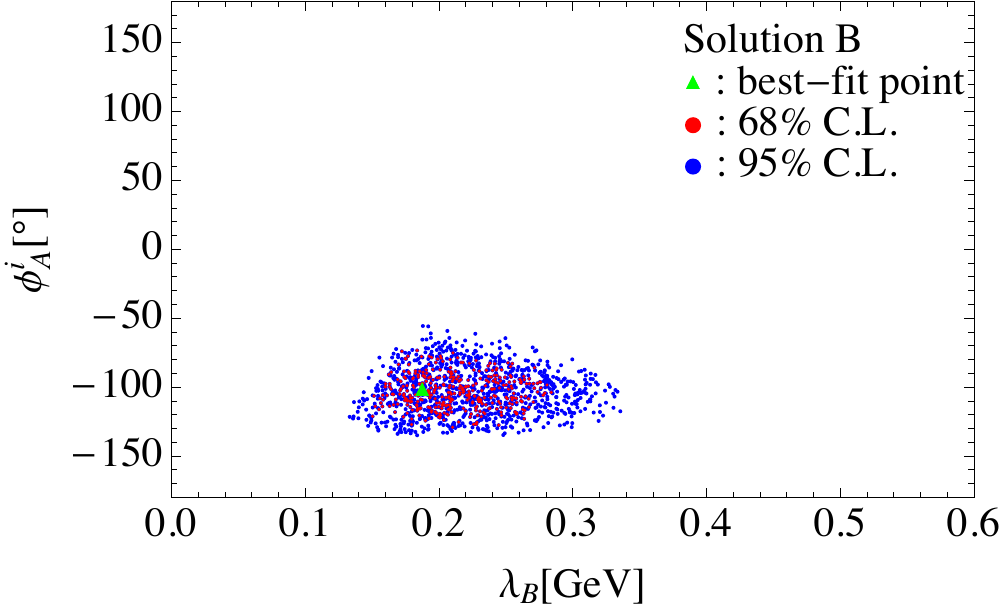}}
 \caption{The fitted results of annihilation parameters
 $\rho_A^{i,f}$, $\phi_A^{i,f}$ and $B$ wave function
 parameter $\lambda_B$ at 68\% C.L. and 95\% C.L..
 The best-fit points of solutions A and B correspond to
 $\chi^2_{\rm min}$ $=$ $4.70$ and $\chi^2_{\rm min}$
 $=$ $4.77$, respectively.}
 \label{finalfit}
 \end{figure}
 \begin{table}[ht]
 \caption{The best-fit values of annihilation
 parameters and wave function parameter $\lambda_B$.}
 \label{finalresult}
 \begin{ruledtabular}
 \begin{tabular}{lccccc}
   & $\rho_{A,H}^i$
   & $\phi_{A,H}^i [^{\circ}]$
   & $\rho_A^f$ & $\phi_A^f[^{\circ}]$
   & $\lambda_B$ [GeV]
   \\ \hline
 Solution A & $2.98^{+1.12}_{-0.86}$ & $-105^{+34}_{-24}$
            & $1.18^{+0.20}_{-0.23}$ & $- 40^{+11}_{-8}$
            & $0.19^{+0.09}_{-0.04}$ \\
 Solution B & $2.97^{+1.19}_{-0.90}$ & $-105^{+32}_{-24}$
            & $2.80^{+0.25}_{-0.21}$ & $165^{+4}_{-3}$
            & $0.19^{+0.10}_{-0.04}$
 \end{tabular}
 \end{ruledtabular}
 \end{table}

 Based on the above analyses and discussions, we present the most simplified
 (flaour conserving) scenario for the annihilation parameters that
 both $(\rho_{A}^{f},\phi_{A}^{f})$ and $(\rho_{A}^{i},\phi_{A}^{i})$
 are universal for both $B_{u,d}$ $\to$ $PP$ and $B_s$ $\to$ $PP$ decay
 modes to lessen phenomenological parameters, 
 where $X^{i}$ and $X^{f}$ are independent variables. 
 To get their exact values, we perform a global fit by combining available
 experimental data for $B_{u,d,s}$ $\to$ $\pi \pi$, $\pi K$, $KK$ decays,
 which involve 16 decay modes and 42 observables.
 In our fit, besides of ($\rho_A^{i,f},\phi_A^{i,f}$), 
 the inverse moment $\lambda_B$, which is used to parameterize integral of 
 the $B$ meson distribution amplitude and a hot topic by now (see Ref.\cite{Beneke5}
 for details), is also treated as a free parameter and taken into account.
 We present the allowed parameter spaces in Fig.\ref{finalfit} and
 the corresponding numerical results in Table \ref{finalresult}.

 \begin{table}[ht]
 \caption{The $CP$-averaged branching ratios (in the unit of $10^{-6}$)
 of $B_{s}$ $\to$ $\pi \pi$, $\pi K$, $KK$ decays, where
 the first and second theoretical errors are caused by uncertainties
 of the CKM and the other parameters
 (including the quark masses, decay constants and form factors) 
 listed in Table \ref{inputs}, respectively.}
 \label{pikbr}
 \begin{ruledtabular}
 \begin{tabular}{lccc}
  Decay mode & Exp. data & This work & Cheng \cite{Cheng2}
  \\ \hline
  $\bar{B}_s \to \pi^- K^+$
  & $5.4 \pm 0.6$ & $5.5^{+0.4+3.4}_{-0.4-2.5}$ & $5.3^{+0.4+0.4}_{-0.8-0.5}$
  \\
  $\bar{B}_s \to \pi^0 K^0$
  & --- & $1.83^{+0.15+0.23}_{-0.16-0.20}$ & $1.7^{+2.5+1.2}_{-0.8-0.5}$
  \\ \hline
  $\bar{B}_s \to \pi^+ \pi^-$
  & $0.73 \pm 0.14$ & $0.61^{+0.02+0.07}_{-0.04-0.06}$ & $0.26^{+0.00+0.10}_{-0.00-0.09}$
  \\
  $\bar{B}_s \to \pi^0 \pi^0$
  & --- & $0.31^{+0.01+0.03}_{-0.02-0.03}$ & $0.13^{+0.0+0.05}_{-0.0-0.05}$
  \\ \hline
  $\bar{B}_s \to K^+ K^-$
  & $24.5 \pm 1.8$ & $20.1^{+0.78+6.1}_{-1.32-5.1}$ &$25.2^{+12.7+12.5}_{-7.2-9.1}$
  \\
  $\bar{B}_s \to K^0 \bar{K}^0$
  & $<66$ & $21.2^{+0.8+6.8}_{-1.4-5.7}$ &$26.1^{+13.5+12.9}_{-8.1-9.4}$
  \end{tabular}
  \end{ruledtabular}
  \end{table}

  \begin{table}[ht]
  \caption{The direct $CP$ asymmetries (in the unit of $10^{-2}$).
  The explanation for uncertainties is the same as in Table \ref{pikbr}.}
  \label{pikdcp}
  \begin{ruledtabular}
  \begin{tabular}{lccc}
  Decay mode & Exp. data & This work & Cheng \cite{Cheng2}
  \\ \hline
  $\bar{B}_s \to \pi^- K^+$
  & $26 \pm 4$ & $31^{+1+14}_{-1-8}$ & $20.7^{+5.0+3.9}_{-3.0-8.8}$
  \\
  $\bar{B}_s \to \pi^0 K^0$
  & --- & $51^{+1+8}_{-2-9}$  & $36.3^{+17.4+26.6}_{-18.2-24.3}$
  \\   \hline
 $\bar{B}_s\to\pi^+\pi^-$            &---               &$0^{+0+0}_{-0-0}$&$0$  \\
 $\bar{B}_s\to\pi^0\pi^0$           &---                  &$0^{+0+0}_{-0-0}$  &$0$ \\
 \hline
  $\bar{B}_s \to K^+ K^-$
  & $-14 \pm 11$ & $-11.6^{+0.4+0.4}_{-0.4-0.4}$ & $-7.7^{+1.6+4.0}_{-1.2-5.1}$
  \\
  $\bar{B}_s \to K^0 \bar{K}^0$
  & --- & $0.54^{+0.02+0.11}_{-0.02-0.13}$  & $0.40^{+0.04+0.10}_{-0.04-0.04}$
 \end{tabular}
 \end{ruledtabular}
 \end{table}

 \begin{table}[ht]
 \caption{The mixing-induced $CP$ asymmetries (in the unit of $10^{-2}$).
  The explanation for uncertainties is the same as in Table \ref{pikbr}.}
  \begin{ruledtabular}
  \label{pikmcp}
  \begin{tabular}{lccc}
  Decay mode & Exp. data & This work & Cheng \cite{Cheng2}
  \\ \hline
  $\bar{B}_s \to \pi^0 K^0$
  & --- & $-10.0^{+4.5+7.0}_{-8.4-7.4}$ & $8^{+29+23}_{-27-26}$
  \\ \hline
  $\bar{B}_s \to \pi^+ \pi^-$
  & --- & $16.4^{+0.6+0.0}_{-0.5-0.0}$ & $15^{+0+0}_{-0-0}$
  \\
  $\bar{B}_s \to \pi^0 \pi^0$
  & --- & $16.4^{+0.6+0.0}_{-0.5-0.0}$ & $15^{+0+0}_{-0-0}$
  \\ \hline
  $\bar{B}_s \to K^+ K^-$
  & $30 \pm 13$ & $18.0^{+0.7+4.3}_{-0.6-5.5}$ & $22^{+4+5}_{-5-3}$
  \\
  $\bar{B}_s \to K^0 \bar{K}^0$
  & --- & $0.50^{+0.02+0.01}_{-0.02-0.02}$  & $0.4^{+0+0.2}_{-0-0.2}$
  \end{tabular}
  \end{ruledtabular}
  \end{table}

 As Fig. \ref{finalfit} shows, the allowed spaces of $(\rho_A^{i,f},\phi_A^{i,f})$
 and $\lambda_B$ are strongly restricted by combined constraints from $B_{u,d,s}$
 $\to$ $\pi \pi$, $\pi K$, $KK$ decays, especially for $(\rho_A^{f},\phi_A^{f})$.
 There are two solutions (named solution A and B, respectively).
 It is easily found that
 the allowed regions and the best-fit point of $(\rho_{A}^{i},\phi_{A}^{i})$
 are so alike that one can hardly distinguish one from these two solutions.
 For each solution, there is no common overlap at 68\% C.L. between
 the allowed regions of $(\rho_{A}^{i},\phi_{A}^{i})$ and
 $(\rho_{A}^{f},\phi_{A}^{f})$ , i.e., the nonfactorizable and factorizable
 annihilation parameters $X_{A}^{i}$ and $X_{A}^{f}$ should be treated
 as independent parameters, which confirms the suggestion of Ref.\cite{zhu2,zhu1}. 
 Numerically, as listed in Table \ref{finalresult}, the fitted result
 is similar to, but with smaller uncertainties, the results in
 Ref.\cite{chang2} where the $B_s$ decay modes are not considered.
 In fact, the two sets of parameters values give the same
 annihilation contributions.
 From Table \ref{finalresult}, it can be seen that a relatively small value 
 of $\lambda_B$ $\sim$ 0.2 GeV which has
 been found by, for instance, Refs. \cite{Beneke2,Cheng2,chang2,Beneke5}
 and a relatively large value of $\rho_{H}$ ${\sim}$ $3$ with 
 ${\phi}_{H}$ ${\sim}$ $-105^{\circ}$ are favored in the 
 phenomenological aspect of $B$ nonleptonic decays. 
 They will enable the HSS corrections to play an important role in evaluating
 observables of penguin dominated $B$ ${\to}$ ${\pi}K$ decays, and 
 have significant enhancement, assisted with the large Wilson coefficient
 $C_{1}$, to the color-suppressed tree amplitude with a large strong
 phase. 
 As noticed and discussed in Refs. \cite{Beneke1,Beneke2,Beneke6}, 
 the vertex corrections, including NLO and NNLO contributions, to the 
 color suppressed tree coefficient $\alpha_2$ exhibit a serious cancellation
 of the real part of $\alpha_{2}$
 (for example, see the first line of Eq.(54) in Ref. \cite{Beneke6}), 
 but the HSS mechanism can compensate for the destructive interference
 and enhance the $\alpha_2$ with a large magnitude. 
 The value of $\alpha_2(\pi\pi)$ ${\simeq}$ $0.24-i\,0.08$ ${\simeq}$ 
 $0.25\,e^{-i\,18^{\circ}}$ including NNLO vertex and HSS corrections
 \cite{Beneke6} obtained with ${\rho}_{H}$ $=$ $0$ and $\lambda_B$ 
 $\sim$ 0.35 GeV still cannot accommodate the experimental data on
 branching ratio $B_{d}$ ${\to}$ ${\pi}^{0}{\pi}^{0}$ decay.  
 So a relatively large HSS corrections arising from $X_{H}$ might be a crucial key
 for the ``$\pi\pi$ puzzle". The branching rate of $B_{d}$ 
 ${\to}$ ${\pi}^{0}{\pi}^{0}$ decay and the $CP$ asymmetry of $B_{u}$
 ${\to}$${\pi}^{0}K^{\pm}$ decay, they both are sensitive to the choice of
 coefficient $\alpha_2$, and can provide substantial constraints on
 parameter $X_{H}$.  
 With the best-fit values of both $(\rho_H,\phi_H)$ and $\lambda_B$ in
 this analysis, one can get $\alpha_2(\pi\pi)$ ${\simeq}$ $0.28-i\,0.49$
 ${\simeq}$ $0.56\,e^{-i\,60^{\circ}}$, which provides a possible solution
 to the so-called ``$\pi\pi$ and $\pi K$" puzzles simultaneously.
 Of course, one can have different mechanism for enhancement of the 
 $\alpha_2$ in QCDF, for example, the final-state rescattering effect
 \footnotemark[2]
 \footnotetext[2]{
 Considering the final state interaction effects, the
 coefficients $\alpha_2(\pi\pi)$ ${\simeq}$ 
 $0.6\,e^{-i\,55^{\circ}}$ ${\simeq}$ $0.34-i\,0.49$ and
 $\alpha_2({\pi}K)$ ${\simeq}$ 
 $0.51\,e^{-i\,58^{\circ}}$ ${\simeq}$ $0.27-i\,0.43$
  \cite{Cheng1}. 
  Notice that (1) the above coefficient $\alpha_2(\pi\pi)$
  has similar magnitude module to ours, and the large
  module of $\alpha_2(\pi\pi)$ is helpful to accommodate
  the ``${\pi}{\pi}$" puzzle.
  (2) The coefficient $\alpha_2({\pi}K)$ has 
  similar magnitude imaginary to ours, and the large imaginary
  part of $\alpha_2({\pi}K)$ results in a large strong phase
  difference to solve the ``${\pi}K$" puzzle. 
  }
 advocated in Ref.\cite{Cheng1} and the Principle of Maximum Conformality
 proposed recently in Ref.\cite{Qiao}, where the allowed regions for 
 parameters ${\rho}_{A,H}$ and ${\phi}_{A,H}$ might be different.
 
  With the inputs in Table \ref{inputs} and the best-fit values of
  parameters listed in Table \ref{finalresult}, we present our
  theoretical results for observables of $B_s$ $\to$ $\pi \pi$,
  $\pi K$, $KK$ decays in the third column of Tables \ref{pikbr},
  \ref{pikdcp} and \ref{pikmcp}.
  The results \cite{Cheng2} with the traditional treatment
  $X_{A}^{i}$ $=$ $X_{A}^{f}$ including flavor symmetry breaking
  effects are also listed in the last column for comparison.
  The results for $B_{u,d}$ $\to$ $\pi \pi$, $\pi K$,  $KK$ decays
  are not listed here, because they are similar to those
  given in Ref.\cite{chang2}.
  From these results, it could be found that (1) all QCDF results
  of $B_{u,d,s}$ $\to$ $\pi \pi$, $\pi K$, $KK$ decays could
  be accommodated to the experimental data within errors.
 (2) Our results of branching ratios for $B_s$ $\to$ $\pi \pi$
 decays are twice as large as those with the traditional treatment
 \cite{Cheng2}.
 And ${\cal B}(B_{s}{\to}{\pi}^{+}{\pi}^{-})$ $=$
 $(0.61^{+0.02+0.07}_{-0.04-0.06}){\times}10^{-6}$ is in 
 good agreement with the data within one experimental error.
 Meanwhile, our result ${\cal B}(B_{s}{\to}{\pi}^{0}{\pi}^{0})$
 $=$ $(0.31^{+0.01+0.03}_{-0.02-0.03}){\times}0^{-6}$ is 
 twice as large as the traditional result 
 $(0.13^{+0.0+0.05}_{-0.0-0.05})\times 10^{-6}$. 
 Moreover, there are also some other differences between
 the two sets of theoretical results more or less. 
 So, the future accurate measurements on the nonleptonic $B_s$
 meson decays would be helpful to probe the annihilation
 contributions and to explore the underlying dynamical
 mechanism.

 In summary, we studied the nonfactorizable and factorizable
 annihilation contributions to $B_{u,d,s}$ $\to$ $\pi \pi$,
 $\pi K$, $KK$ decays with QCDF approach.
 To clarify the independence of annihilation parameters $X_{A}^{i}$
 and $X_{A}^{f}$ and the possible flavor symmetry breaking effects
 therein, a statistical $\chi^2$ analysis is performed for nonleptonic
 $B_{u,d}$ and $B_{s}$ decays.
 It is found that (1) $X_{A}^{i}$ and $X_{A}^{f}$ are independent parameters,
 which differs from the traditional treatment with annihilation parameters 
 and verifies the proposal of Ref.\cite{zhu2}.
  (2) The flavor symmetry breaking effects might be small for nonleptonic 
  $B_{u,d}$ and $B_{s}$ decays by now due to the large experimental 
  errors and theoretical uncertainties.
 With the simplifications $X_{A,s}^i$ $=$ $X_{A,d}^i$ and
 $X_{A,s}^f$ $=$ $X_{A,d}^f$, a comprehensive global fit
 on the annihilation parameters and the $B$ wave function
 parameter $\lambda_B$ is done based on the current available
 measurements on $B_{u,d,s}$ $\to$ $\pi \pi$, $\pi K$, $KK$
 decays. Two allowed solutions are found.
 With the best-fit parameters summarized in Table \ref{finalresult},
 the QCDF results for $B$ $\to$ $\pi \pi$, $\pi K$, $KK$ decays
 are consistent with the present experimental data within errors.
 It is expected that the measuremental precision of nonleptonic
 $B$ decays could be much improved by LHCb and super-B experiments
 in the following years, so more information about annihilation
 contributions could be revealed.

 \section*{Acknowledgments}
 The work is supported by the National Natural Science Foundation of China
 (Grant Nos. 11105043, 11147008, 11275057, 11475055 and U1232101).
 Q. Chang is also supported by the Foundation for the Author of National
 Excellent Doctoral Dissertation of P. R. China (Grant No. 201317)
 and the Program for Science and Technology Innovation Talents in
 Universities of Henan Province (Grant No. 14HASTIT036).
 Thanks Referees for their helpful comments 
  and Xinqiang Li for helpful discussions.

 \end{document}